\documentclass[prb,superscriptaddress,floatfix,twocolumn,showpacs,amsfonts,amsmath,amssymb]{revtex4}
\usepackage{graphicx} 
\usepackage{dcolumn} 
\usepackage{bm} 
\usepackage{color}

\begin{document}

\title{Magnetic Excitations in the Spin-Spiral State of TbMnO$_3$
and DyMnO$_3$}

\author{Alexander I. Milstein}
\affiliation{Budker Institute of Nuclear Physics, 630090 Novosibirsk, Russia}
\author{Oleg P. Sushkov}
\affiliation{School of Physics, University of New South Wales, Sydney 2052, Australia}
\affiliation{Yukawa Institute for Theoretical Physics, Kyoto University,
Kyoto 606-8317, Japan}

\date{\today}

\begin{abstract}
We calculate spectra of magnetic excitations in the spin-spiral state of 
perovskite manganates. The spectra consist of several branches
corresponding to different polarizations and different ways of diffraction
from the static magnetic order.
Goldstone modes and opening of gaps at zero and
non-zero energies due to the crystal field and the 
Dzyaloshinski-Moriya anisotropies are discussed. 
Comparing results of the calculation with available experimental data
we determine values of effective exchange parameters and  anisotropies.
To simplify the spin-wave calculation and to get a more clear physical
insight in the structure of excitations we use the $\sigma$-model-like
effective field theory to analyze the Heisenberg Hamiltonian and to
derive the spectra.
\end{abstract}

\pacs{
74.72.Dn, 
75.10.Jm, 
75.50.Ee 
}

\maketitle

\section{Introduction}
Terbium and Dysprosium manganates,  TbMnO$_3$ and DyMnO$_3$, are the key
materials in the family of multiferroic oxides~\cite{Kimura2003,Goto2004}.
Properties of TbMnO$_3$ and DyMnO$_3$ are very similar, to be specific
below we consider  TbMnO$_3$.
Similar to the parent compound of the rare-earth manganites LaMnO$_3$, 
TbMnO$_3$ has orthorhombic lattice structure with lattice constants 
$a \approx 5.302 \AA$, $b = 5.857\AA$, and $c = 7.402\AA$, 
Ref. \cite{Blasco2000} 
Below we measure components of wave vectors in units $1/a$, $1/b$, and $1/c$ accordingly.
There are three different magnetic phase transitions 
in  TbMnO$_3$ upon cooling \cite{Kajimoto2004,Kenzelmann2005}. 
An incommensurate collinear spin-density wave
with  the wave vector directed along {\bf b}, ${\bm Q}\approx \pi(0,0.28,0)$,
and Mn spins also aligned along {\bf b} is stabilized below $T_N=42$K.
This is the spin-stripe phase which is also called the ``sinusoidal
phase''.
Below $T_S=28K$ Mn spins reorient into an incommensurate spin spiral. The wave vector of the spiral
is practically the same as that in the spin stripe phase, Mn spins are confined in the {\bf bc}-plane.
Finally  Tb spins order below $T=7$K.
Last but not least, simultaneously with the transition into the spin-spiral phase an electric polarization along {\bf c} appears  at $T=T_S$~\cite{Kimura2003}.
The polarization is coupled with the spin-spiral 
due to  the Dzyaloshinski-Moriya interaction \cite{Katsura2005,Mostovoy2006}.

In the present work we concentrate on magnetic properties and do not consider ferroelectricity.
The major magnetic properties are related to Mn ions.
On the other hand Tb ions, which order at the relatively low temperature, 
play a minor role. In our analysis we disregard Tb ions.
There are two very important points concerning magnetic properties of
the rare-earth manganites: (i) Magnetic excitations in the spin-spiral phase
measured in Ref.~\cite{Senff2008} are quite unusual.
(ii) Even more unusual is the spin-spiral to spin-stripe phase transition at 
$T=T_s$. 
The phase transition has been considered phenomenologically within an effective
Landau-Ginzburg theory in Ref.~\cite{Mostovoy2006}.
We believe that physics behind points (i) and (ii) are closely related,
the unusual excitation spectrum is behind the unusual phase transition.
In the present paper we address only the first point, we calculate magnetic
excitations in the spin-spiral phase.
A brute force spin-wave calculation of excitations in the spin-spiral phase
is certainly possible, but it is rather technically involved.
More importantly such a calculation is  not  transparent physically.
Because of this reason we employ a much more transparent/efficient
 $\sigma$-model like field theory to find excitations. A similar approach 
was used previously for calculation of magnetic  excitations in the spin-spiral 
compounds  FeSrO$_3$ and FeCaO$_3$~\cite{Milstein2011}.
The field theory  is well justified at small momenta,
while close to the boundary of magnetic Brillouin zone it can have up to
20-30\% inaccuracy. We sacrifice this to get a transparent description
of the most important incommensurate physics at small momenta.
Structure of the paper is the following:
In Section II we consider collinear antiferromagnet
LaMnO$_3$ and formulate the field theory.
In this case the spin-wave calculation is straightforward and we compare it 
with the field theory.
In Section III we calculate magnetic excitations in the spin-spiral phase
without account of anisotropies and discuss Goldstone modes.
Influence of the single ion anisotropy on excitation spectra
is considered in Section IV.
In Section V we consider the combined influence of 
the single ion anisotropy and the Dzyaloshinski-Moriya
anisotropy on excitation spectra.
All the plots in Sections III, IV, and V are presented at values of parameters
which reproduce the experimental spectra from Ref.~\cite{Senff2008}.
Those readers who are not interested in details of the calculations can
go directly to Section VI where we summarize the results, 
refer to plots showing the calculated dispersions, and present our
conclusions.

\section{Spin-wave and field theory calculations of magnetic excitations
in ${\mbox {LaMnO}}_3$}
\begin{figure}[ht]
\includegraphics[width=0.3\textwidth,clip]{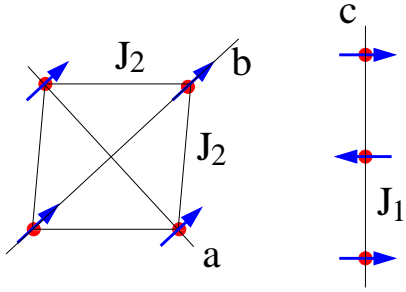}
\caption{Magnetic structure of LaMnO$_3$, ferromagnetic ordering in the ab-plane and
antiferromagnetic ordering along the c-axis.
}
\label{F1}
\end{figure}

Magnetic structure as well as magnetic excitations in LaMnO$_3$ have been
determined by neutron scattering~\cite{Moussa1996,Hirota1996}.
In the ab-plane spins of Mn ions are aligned ferromagnetically, while in the
c-direction they are aligned antiferromagnetically, Fig.\ref{F1}.
The minimal Heisenberg Hamiltonian describing the system 
is~\cite{Moussa1996,Hirota1996}
\begin{equation}
\label{h1}
H=J_1\sum_{\langle i,j\rangle_c}{\vec S}_i\cdot{\vec S}_j 
-J_2\sum_{\langle i,j\rangle_{ab}}{\vec S}_i\cdot{\vec S}_j \ ,
\end{equation}
where $S=2$ is spin of Mn ion, $\langle i,j\rangle_c$ denotes nearest neighbours in the c-direction
and $\langle i,j\rangle_{ab}$ denotes nearest neighbours in the ab-plane,
$J_1$ and $J_2$ are antiferromagnetic and ferromagnetic exchange integrals indicated
in Fig.\ref{F1}.
In this work we use the standard definition of exchange integrals: each link in (\ref{h1})
is counted only once. Therefore, our exchange integrals are twice larger than that defined in 
Refs.\cite{Senff2008,Moussa1996,Hirota1996}.
We do not account in (\ref{h1}) the single ion anisotropy because the goal of the present section
is just to introduce the field theory.
The spin-wave diagonalization of the Hamiltonian (\ref{h1}) is straightforward (a combination
of Holstein-Primakoff and Bogoliubov's transforms). This results in the following magnon 
dispersion \cite{Moussa1996,Hirota1996}
\begin{widetext}
\begin{eqnarray}
\label{sw}
&&A_{\bm q}=J_1+2J_2(1-\cos q_a cos q_b)\nonumber\\
&&B_{\bm q}=J_1\cos q_c\nonumber\\
&&\omega_{\bm q}=2S\sqrt{A_{\bm q}^2-B_{\bm q}^2}=2S\sqrt{J_1^2\sin^2q_c+4J_1J_2(1-\cos q_a cos q_b)+4J_2^2(1-\cos q_a cos q_b)^2}
\end{eqnarray}
\end{widetext}

It is well known that in the long wave-length limit, $q\ll \pi$,
any quantum antiferromagnet is equivalent to a non-linear $\sigma$-model
written in terms of the unit vector ${\vec n}$ describing the 
staggered magnetization. The effective Lagrangian of the $\sigma$-model
reads
\begin{equation}
\label{L}
{\cal L}=\frac{1}{2}\chi_{\perp}{\dot {\vec n}}^2-E({\vec n})\ ,
\end{equation}
where $\chi_{\perp}$ is perpendicular magnetic susceptibility
and $E({\vec n})$ is energy of elastic deformation of spin fabric.
\begin{figure}[ht]
\includegraphics[width=0.1\textwidth,clip]{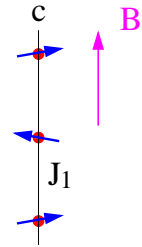}
\caption{Response of spins to the magnetic field ${\vec B}$
applied perpendicular to the staggered magnetization.
}
\label{F2}
\end{figure}
The magnetic susceptibility corresponds to the interaction Hamiltonian
$H_B=-\sum_i{\vec B}\cdot{\vec S}_i$, with magnetic field ${\vec B}$
applied perpendicular to the staggered magnetization, see
Fig.\ref{F2}. A simple calculation shows that the susceptibility per site is
\begin{equation}
\label{chi}
\chi_{\perp}=\frac{1}{4J_1} \ .
\end{equation}
The elastic energy corresponding to the Hamiltonian (\ref{h1}) is
\begin{eqnarray}
\label{E}
&&E=-S^2{\vec n}R_{0}(\bm p){\vec n}+const \ ,\nonumber\\
&&R_{0}=\frac{J_1}{4}\cos(2p_c) +2J_2\cos p_a\cos p_b \nonumber\\
&&p_a=-i\nabla_a \ , \ \ p_b=-i\nabla_b \ , \ \ p_c=-i\nabla_c \ .
\end{eqnarray}
Usually $E$ is expanded up to the second order in momentum,
$E \to {\vec n}\left\{\frac{\rho_{ab}}{2}(p_a^2+p_b^2)
+\frac{\rho_{c}}{2}p_c^2\right\}{\vec n}\to
\frac{\rho_{ab}}{2}[(\nabla_a{\vec n})^2+(\nabla_b{\vec n})^2]
+\frac{\rho_{c}}{2}(\nabla_c{\vec n})^2$, where $\rho_{ab}$ and $\rho_c$
are the corresponding spin stiffnesses.
In the present work we do not expand $E$ in powers of momentum, instead
we use (\ref{E}) as it is. Note that the ferromagnetic $J_2$-term in
(\ref{E}) is unambiguous, on the other hand the antiferromagnetic 
$J_1$-term is somewhat ambiguous. One can write  the antiferromagnetic 
$J_1$-term  as it is done in (\ref{E}) or alternatively as $J_1\cos(p_c)$.
In the long-wave length limit the both ways result in the same spin
stiffness $\frac{J_1}{4}\cos(2p_c)\to const-J_1p_c^2/2$, and
$J_1\cos(p_c)\to const-J_1p_c^2/2$. We use the way (\ref{E}) because it
leads to the correct magnon dispersion up to $p_c=\pi/2$, see Eq.(\ref{oq}),
and hence allows one to overstretch the region of validity of the
field theory~\cite{com1}.

Minimum of energy (\ref{E}) defines the ground state which corresponds
to the constant staggered magnetization ${\vec n}={\vec n}_0$.
Magnetic excitations above the ground state, ${\vec n}={\vec n}_0+
\delta{\vec n}$, $\delta{\vec n} \perp {\vec n}_0$,
are defined by the Euler-Lagrange equation of Lagrangian (\ref{L}).
\begin{equation}
\label{EL1}
\chi_{\perp}{\ddot{\delta{\vec n}}}=
2S^2\left[R_{0}({\bm p})-R_{0}(0)\right]\delta{\vec n}\ .
\end{equation}
For $\delta{\vec n}=\delta{\vec n}_0e^{-i\omega_{\bm q}t+i{\bm q}\cdot{\bm r}}$ this
results in the dispersion
\begin{equation}
\label{oq}
\omega_{\bm q}=2S\sqrt{J_1^2\sin^2q_c+4J_1J_2(1-\cos q_a cos q_b)} \ .
\end{equation}
Compared to the ``exact'' spin-wave calculation (\ref{sw}) the
term $4J_2^2(1-\cos q_a cos q_b)^2$ is missing under the square root.
In the long wave-length limit, $q_a,q_b \ll \pi$, this term is quartic in
momenta and therefore it is irrelevant. Moreover, at $J_2 \ll J_1$
this term is irrelevant even at $q_a,q_b = \pi$.
The inequality $J_2 \ll J_1$ is certainly not valid for
LaMnO$_3$ where $J_1\approx 1.17$ meV and $J_2\approx 1.66$ meV,
see Refs.~\cite{Moussa1996,Hirota1996}.
However, we will see that for TbMnO$_3$ $J_2 \lesssim J_1/2$.

For the  collinear magnetic ground state in LaMnO$_3$ the spin-wave
calculation (\ref{sw}) is very simple and therefore application
of the field theory does not make sense. The purpose of the present section
is just to demonstrate how the field theory works in the known simple case.
Below we employ the field theory for the spin-spiral states of TbMnO$_3$ 
and DyMnO$_3$. For a noncollinear state the field theory is significantly
more  efficient technically.

It is instructive to compare also quantum/thermal fluctuations obtained
within the spin-wave theory and within the field theory.
The fluctuation reduction of the staggered magnetization within the 
spin-wave theory is
determined by Bogoliubov's parameters $u_{\bm q}$ and $v_{\bm q}$:
\begin{eqnarray}
\label{bog}
&&u_{\bm q}^2=\frac{1}{2}\left(A_{\bm q}/\sqrt{A_{\bm q}^2-B_{\bm q}^2}+1\right)
\nonumber\\
&&v_{\bm q}^2=\frac{1}{2}\left(A_{\bm q}/\sqrt{A_{\bm q}^2-B_{\bm q}^2}-1\right)\nonumber\\
&&\langle n_b\rangle=\frac{\langle S_b\rangle}{S}=
1-\frac{2}{S}\sum_{q\in MBZ}\left\{v_{\bm q}^2+(u_{\bm q}^2+v_{\bm q}^2)f_{\bm q}\right\}
\nonumber\\
&&=1-\sum_{q\in MBZ}\left\{\left(\frac{2A_{\bm q}}{\omega_{\bm q}}-\frac{2}{S}\right)
+\frac{4A_{\bm q}}{\omega_{\bm q}}f_{\bm q}\right\} \ .
\end{eqnarray}
Here $f_{\bm q}=\left(e^{\omega_{\bm q}/T}-1\right)^{-1}$ is the Bose thermal occupation
factor. The summation over momentum is performed inside the Magnetic 
Brillouin Zone (MBZ), 
$-\pi/2 \leq q_c \leq \pi/2$, $-\pi \leq q_a,q_b \leq \pi$.
The fluctuation reduction within the field theory is
of the following form~\cite{Milstein2011}
\begin{eqnarray}
\label{rfi}
&&\langle n_b\rangle=
1-\sum_{q \in MBZ}\frac{1}{\chi_{\perp}\omega_{\bm q}}
\left(\frac{1}{2}+f_{\bm q}\right) \ .
\end{eqnarray}
At small $q$ the integrand in (\ref{rfi}) is equal to that in
(\ref{bog}), this is true for both thermal fluctuations
(proportional to $f_{\bf q}$) and for quantum fluctuations.
Moreover, at $J_2 \ll J_1$ the thermal fluctuation contributions 
in Eqs. (\ref{rfi}) and (\ref{bog}) are equal
over the entire MBZ. The large $q$ quantum fluctuation contributions
in Eqs. (\ref{rfi}) and (\ref{bog}) are generally
different. However, for S=2 quantum fluctuations are anyway small and there 
is no need to consider them.

\section{Magnetic excitations in the spin-spiral phase of 
$\mbox{TbMnO}_3$
without account of anisotropies}
According to Ref.~\cite{Senff2008} the incommensurate spin structure in 
TbMnO$_3$ is due to ab-plane frustrating antiferromagnetic interaction $J_{3b}$
\begin{figure}[ht]
\includegraphics[width=0.15\textwidth,clip]{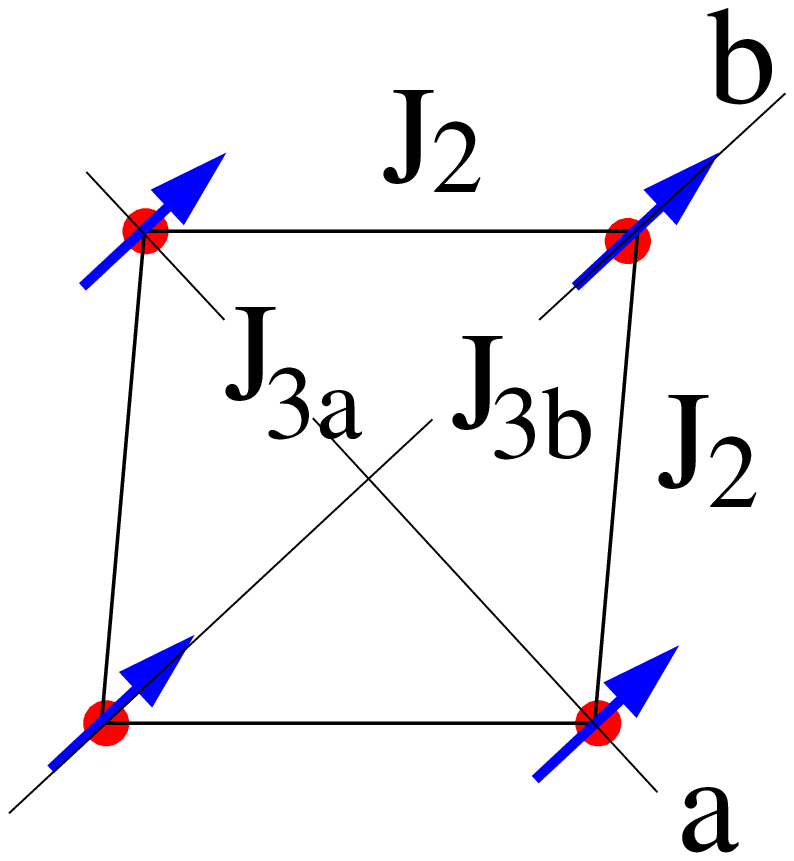}
\caption{Frustrating ab-plane antiferromagnetic interactions $J_{3b}$
and $J_{3a}$ in TbMnO$_3$.}
\label{F3}
\end{figure}
shown in Fig.\ref{F3}, for completeness we also introduce $J_{3a}$.
So,  in TbMnO$_3$ there is the following addition
to the Hamiltonian (\ref{h1})
\begin{equation}
\label{h2}
\delta H=J_{3b}\sum_{\langle i,j\rangle_b}{\vec S}_i\cdot{\vec S}_j 
+J_{3a}\sum_{\langle i,j\rangle_{a}}{\vec S}_i\cdot{\vec S}_j \ .
\end{equation}
Here $\langle i,j\rangle_b$ denotes next nearest neighbours along 
the b-direction and $\langle i,j\rangle_{a}$ denotes the next nearest 
neighbours along the a-direction.
The  spin-elastic energy corresponding to $H+\delta H$
is similar to (\ref{E})
\begin{eqnarray}
\label{E1}
E&=&-S^2{\vec n}R(\bm p){\vec n}+const \ ,\nonumber\\
R&=&\frac{J_1}{4}\cos(2p_c) +2J_2\cos p_a\cos p_b  \nonumber\\
 &-&J_{3b}\cos(2p_b) -J_{3a}\cos(2p_a) \ .
\end{eqnarray}
Below we assume that 
\begin{equation}
\label{ineq}
J_2<2J_{3b}, \ \ \ \  J_2^2>4J_{3a}J_{3b} \ .
\end{equation}
In this case it is easy to check that the energy (\ref{E1})
is minimum  for the  spin spiral ground state
\begin{eqnarray}
\label{n0}
&&{\vec n}_0={\vec e}_1 \cos(\bm Q\cdot\bm r)+{\vec e}_2 \sin(\bm Q\cdot\bm r)\,,\nonumber\\
&& \bm Q= Q \bm e_b\,,\quad \cos Q=\frac{J_2}{2J_{3b}}\,,
 \end{eqnarray}
where ${\vec e}_1$ and ${\vec e}_2$ are two arbitrary orthogonal unit vectors
which define plane of the spiral.
According to Ref.~\cite{Senff2008} in TbMnO$_3$ the wave vector is
$Q \approx 0.28\pi$, hence  $J_2/J_{3b}\approx 1.27$.\\

{\underline {\large In-plane excitations.}}\\
There are two types of magnetic excitations in the spin spiral state, 
in-plane spin excitation and out-of-plane spin excitation. The in-plane 
excitation is described by a phase $\varphi(t,{\bm r})$,
$\varphi\ll 1$, it results in the following vector ${\vec n}$,
\begin{eqnarray}
\label{nin}
&&{\vec n}={\vec e}_1 \cos(\bm Q\cdot\bm r+\varphi)+{\vec e}_2 \sin(\bm Q\cdot\bm r+\varphi)\,,\nonumber\\
&&\approx\left(1-\varphi^2/2\right){\vec n}_0+\varphi{\vec n}_1 \,,\nonumber\\
&&{\vec n}_1=-{\vec e}_1 \sin(\bm Q\cdot\bm r)+{\vec e}_2 \cos(\bm Q\cdot\bm r)\,.
 \end{eqnarray}
Substituting this ${\vec n}$ in Eqs.(\ref{L}),(\ref{E1}) and taking
variation with respect to $\varphi$  we find the following Euler-Lagrange 
equation
\begin{eqnarray}
\label{eqphi}
-\chi_{\perp}{\ddot {\varphi}}
+2S^2[-\varphi{\vec n}_0 R(\bm p){\vec n}_0+{\vec n}_1 R(\bm p){\vec n}_1\varphi]=0 \,.
 \end{eqnarray}
Having in mind the plane wave solution, $\varphi=\varphi_0\exp(-i\omega_{\bm q}t+i\bm q\cdot\bm r)$,
we note that the following relations are valid
\begin{eqnarray}
\label{rel}
&&{\vec n}_1 R(\bm p){\vec n}_1e^{i\bm q\cdot\bm r}=
\frac{1}{2}[R(\bm q+\bm Q)+R(\bm q-\bm Q)]e^{i\bm q\cdot\bm r}\nonumber\\
&&{\vec n}_0 R(\bm p){\vec n}_0=R(\bm Q)\ .
\end{eqnarray}
Hence Eq.(\ref{eqphi}) results in the following spectrum of the in-plane excitation
\begin{eqnarray}
\label{omegain}
\omega_{\bm q}^{(in)}=2S\sqrt{J_1[2R(\bm Q)-R(\bm q+\bm Q)-R(\bm q-\bm Q)]} \,.
 \end{eqnarray}
As one should expect $\omega_{\bm q}^{(in)}=0$ for $\bm q=0$. This is the
Goldstone sliding mode.\\

{\underline {\large Out-of-plane excitations.}}\\
The out-of-plane excitation $h(t,{\bm r})$,
$h \ll 1$, results in the following vector ${\vec n}$,
\begin{eqnarray}
\label{nout}
{\vec n}=\sqrt{1-h^2}\,{\vec n}_0+h{\vec e}_3  
\approx\left(1-h^2/2\right){\vec n}_0+h{\vec e}_3  \,,
 \end{eqnarray}
where  ${\vec e}_3=[{\vec e}_1\times {\vec e}_2]$ is a unit vector 
perpendicular to the plane of spiral. Substituting (\ref{nout})
in Eqs.(\ref{L}),(\ref{E1}) and performing variation with respect to $h$, 
we get the following Euler-Lagrange equation
\begin{eqnarray}
\label{eqh}
-\chi_{\perp}{\ddot h}
+2S^2[-h{\vec n}_0 R(\bm p){\vec n}_0+{\vec e}_3 R(\bm p){\vec e}_3h]=0 \,.
 \end{eqnarray}
The plane-wave solution, $h=h_0\exp(-i\omega_{\bm q}t+i\bm q\cdot\bm r)$,
gives the following spectrum of the out-of-plane excitation
\begin{eqnarray}
\label{omegaout}
\omega_{\bm q}^{(out)}=2S\sqrt{2J_1[R(\bm Q)-R(\bm q)]} \,.
 \end{eqnarray}
The dispersion has two zeroes (Goldstone modes) $\omega_{\bm q}^{(out)}=0$ for $\bm q=\pm{\bm Q}$. 

Altogether the spectrum has three Goldstone modes corresponding
to three possible rotations of the spin spiral. The in-plane
sliding mode with $q=0$ corresponds to the rotation around ${\vec e}_3$,
and two out-of-plane modes with $\bm q=\pm{\bm Q}$ correspond to linear
combinations of rotations around ${\vec e}_1$ and ${\vec e}_2$. \\

{\underline {\large Comparison with experiment.}}\\
Dispersions of two branches (\ref{omegain}) and (\ref{omegaout}) have been
derived without account of anisotropies. The anisotropies, which we consider
later, significantly modify the dispersions at small momenta.
However, close to boundaries of  MBZ, where excitation energies are
sufficiently high, influence of anisotropies is relatively small.
Therefore, to estimate values of the exchange integrals we calculate
$\omega_{\bm q}^{(out)}$ at some points at the boundary of MBZ.
According to Eq.(\ref{omegaout})
\begin{eqnarray}
\label{MBZq}
&&{\bm q}=(0,Q,\frac{\pi}{2}),\ \ \ \omega_{\bm q}^{(out)}=2SJ_1\nonumber\\
&&{\bm q}=(\pi,Q,0),\ \ \ \ \omega_{\bm q}^{(out)}=5.1S\sqrt{J_1J_{3b}}\nonumber\\
&&{\bm q}=(0,\pi,0),\ \ \ \ \ \omega_{\bm q}^{(out)}=6.5S\sqrt{J_1J_{3b}}
\end{eqnarray}
Comparing this with data presented in Figs.8,10 from Ref.~\cite{Senff2008} 
we find approximate values of the exchange integrals
\begin{eqnarray}
&&J_1\approx 0.9\,\mbox{meV}\,,\, J_{2}\approx 0.38 \,\mbox{meV}\nonumber\\
&&J_{3b}\approx0.3 \,\mbox{meV}\,,\, J_{3a}=0.1\,\mbox{meV}\,.
\label{parameters}
\end{eqnarray}
Note that $J_2$ follows from Eq.(\ref{n0}) as soon as $J_{3b}$ is
determined. There are no data to determine $J_{3a}$. Rather arbitrarily
we take $J_{3a}=0.1$meV which satisfies the inequality (\ref{ineq}).
Values of $J_2$, $J_{3b}$, and $J_{3a}$ presented in (\ref{parameters})
are probably slightly larger than the real ones ($\sim$20\%) because of the 
inaccuracy of the  field theory close to the boundary of MBZ.
Values of exchange integrals in Eq. (\ref{parameters}) reasonably agree
with that derived in Ref.~\cite{Senff2008} (We remind that our
integrals are formally by factor 2 larger due to the different definition).

\begin{figure}[h!]
\includegraphics[width=0.25\textwidth,clip]{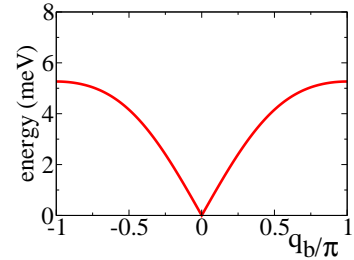}\\
\caption{The in-plane magnon dispersion without account of anisotropies.
The dispersion is shown along the ${\bm q}=(0,q_b,0)$ direction.
} 
\label{F4a}
\end{figure}
\begin{figure}[h!]
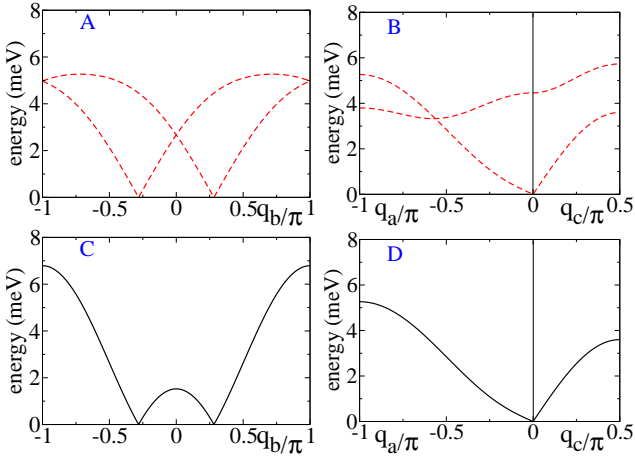

\includegraphics[width=0.23\textwidth,clip]{Fig4bI.eps}
\includegraphics[width=0.23\textwidth,clip]{Fig4cI.eps}
\includegraphics[width=0.23\textwidth,clip]{Fig4b.eps}
\includegraphics[width=0.23\textwidth,clip]{Fig4c.eps}
\caption{Magnon dispersions (as they are seen in neutron scattering)
without account of anisotropies.
Panels A and B: branches of the in-plane dispersion 
$\omega^{in}({\bm q}\pm {\bm Q})$.
Panels C and D: out-of-plane dispersion $\omega^{out}({\bm q})$.
Panels A and C correspond to {\bf q} along {\bf b}, ${\bm q}=(0,q_b,0)$.
Panels B and D correspond to ${\bm q}={\bm Q}+\delta{\bm q}$, where
$\delta{\bm q}$ is directed along {\bf a} and {\bf c}.
} 
\label{F4}
\end{figure}

The in-plane dispersion (\ref{omegain}) has minimum at $q=0$. The dispersion 
for ${\bm q}=(0,q_b,0)$ is shown in Fig.\ref{F4a}.
The in-plane excitation shown in Fig.\ref{F4a} cannot be seen directly in 
neutron scattering since the corresponding n-field (\ref{nin}) contains an 
additional oscillating factor 
$\cos(\bm Q\cdot\bm r)$ or $\sin(\bm Q\cdot\bm r)$. Therefore in a scattering 
measurement the in-plane mode is seen as two shifted branches 
$\omega^{in}({\bm q}\pm{\bm Q})$ with half intensity each.
These branches are shown by red dashed lines in Fig.\ref{F4}, panels A and B,
along three different directions. Note, there is a crossing in panel B
 at $q_a=\pm 2Q$.
The out-of-plane excitation (\ref{nout}),(\ref{omegaout}) can be seen
in inelastic neutron scattering as it is. The corresponding dispersion
(\ref{omegaout}) along three different directions is plotted in Fig.\ref{F4},
panels C and D, by black solid lines. 

\section{Excitation spectra with account for the crystal field anisotropy along the ${\mbox b}$-axis}
Different anisotropies influence the magnon spectra in different ways.
In this section we consider {\it only} the crystal field anisotropy 
along the b-axis.
The corresponding correction to the elastic energy (\ref{E1}) is
\begin{equation}
\label{db}
H_{cf}=-\Lambda S_b^2=-\Lambda S^2 n_b^2\ ,
\end{equation}
where $\Lambda > 0$ is the strength of the crystal field.
While in the present work we consider only the
spin-spiral phase, the sign of $\Lambda$ (``easy axis'' anisotropy) is dictated
by the spin-stripe phase, where spin is directed along {\bf b}.
We assume that $\Lambda$ is sufficiently small and therefore consider
only effects linear in $\Lambda$.
Dispersion plots presented below correspond to
\begin{equation}
\label{gbb}
\Lambda=0.125 \ {\mbox {meV}}\ ,
\end{equation}
which is approximately consistent with the neutron scattering 
data~\cite{Senff2008}.
The crystal field (\ref{db}) results in two static effects. (i)
The plane of the spin spiral  must include the axis {\bf b}. So, while in 
Eq.(\ref{n0}) vectors ${\bm e}_1$ and ${\bm e}_2$ are arbitrary orthogonal
unit vectors, now we take 
\begin{equation}
\label{eb1}
{\bm e}_2={\bm e}_b\ , \ \ \  {\bm e}_1 \perp {\bm e}_b\ ;
\end{equation}
(ii) The spin spiral gets an additional static position dependent phase
$\varphi_{st}({\bm r})$. So (\ref{n0}) is replaced by
\begin{eqnarray}
\label{N0}
{\vec N}_0&=&{\vec e}_1 \cos(\bm Q\cdot\bm r+\varphi_{st})+{\vec e}_2 \sin(\bm Q\cdot\bm r+\varphi_{st})\nonumber\\
&=&\cos\varphi_{st}\,{\vec n}_0+\sin\varphi_{st}\,{\vec n}_1 \,.
\end{eqnarray}
\\

{\underline {\large Static deformation of the spin spiral.}}\\
Minimization of energy $E+H_{cf}$, Eqs. (\ref{E1}),(\ref{db}),
results in the following equation for
$\varphi_{st}$
\begin{eqnarray}
\label{eqphist}
[-\varphi_{st}{\vec n}_0 R(\bm p){\vec n}_0+{\vec n}_1 R(\bm p){\vec n}_1\varphi_{st}]= -\Lambda\sin(2\bm Q\cdot \bm r) \,.
 \end{eqnarray}
Solution of this equation is
\begin{eqnarray}
\label{phist}
\varphi_{st}({\bm r})=-\frac{\Lambda\sin(2\bm Q\cdot \bm r)}{R(3\bm Q)-R(\bm Q)}=
\frac{\Lambda\sin(2\bm Q\cdot \bm r)}{8J_{3b}\sin^2(2Q)\sin^2Q} \,.
 \end{eqnarray}
The phase of the spin spiral  (\ref{N0}) is $\Phi=\bm Q\cdot\bm r+\varphi_{st}$. 
The phase has a zero mode corresponding to the shift ${\bm r}\to 
{\bm r}+\delta{\bm r}$ 
\begin{equation}
\label{zerophi}
\varphi({\bm r}) \propto \frac{\partial \Phi({\bm r})}
{\partial({\bm Q}\cdot{\bm r})}
=\left\{1-
\frac{2\Lambda \cos(2\bm Q\cdot\bm r)}{R(3\bm Q)-R(\bm Q)}\right\}\,.
\end{equation}
This is the Goldstone sliding mode which remains gapless 
 in presence of anisotropy, $\omega_{q=0}^{(in)}=0$. \\

{\underline {\large In-plane excitations.}}\\
According to the discussion in the previous paragraph, the in-plane
excitation remains gapless even with the anisotropy.
The only qualitatively visible effect of the anisotropy  is 
discontinuity of the dispersion due to diffraction of magnons
from the static spin spiral. The dispersion is discontinuous 
at $q_b=Q$ and $q_b=\pi- Q$.

To find the  in-plane excitation with nonzero energy
$\varphi(t,{\bm r})$ we represent the vector ${\vec n}$
similar to (\ref{nin})
\begin{eqnarray}
\label{ninnew}
&&{\vec n}={\vec e}_1 \cos(\bm Q\cdot\bm r+\varphi_{st}+\varphi)
+{\vec e}_2 \sin(\bm Q\cdot\bm r+\varphi_{st}+\varphi)\,,\nonumber\\
&&\approx\left(1-\frac{\varphi^2}{2}\right){\vec N}_0+\varphi{\vec N}_1 \,,\nonumber\\
&&{\vec N}_1=-{\vec e}_1 \sin(\bm Q\cdot\bm r+\varphi_{st} )+{\vec e}_2 \cos(\bm Q\cdot\bm r+\varphi_{st} )\,,
 \end{eqnarray}
The corresponding Euler-Lagrange equation is
\begin{eqnarray}
\label{eqphinew}
&&-\chi_{\perp}{\ddot {\varphi}}
+2S^2[-\varphi{\vec N}_0 R(\bm p){\vec N}_0+{\vec N}_1 R(\bm p){\vec N}_1\varphi]
\nonumber\\
&& + 2\Lambda S^2\varphi\cos(2\bm Q\cdot\bm r)=0 \,.
 \end{eqnarray}
It is easy to check that the zero frequency sliding mode solution 
(\ref{zerophi})  satisfies this equation.

The spin spiral in combination with the crystal field anisotropy
(\ref{db}) generates the effective scattering ``potential'' with momentum 
$\Delta {\bm q}=2{\bm Q}$.
As usual, the scattering is most pronounced when the ``resonance'' condition,
$\omega_{\bm q}^{(in)}=\omega_{\bm q\pm 2{\bm Q}}^{(in)}$, is fulfilled. The condition
is fulfilled at ${\bm q}={\bm q}_{\perp}+{\bm Q}=(q_a,Q,q_c)$ and at
${\bm q}={\bm q}_{\perp}+{\bm \pi}_b-{\bm Q}=(q_a,\pi-Q,q_c)$.
At these planes the magnon spectrum becomes discontinuous.
Eq.(\ref{eqphinew}), which describes magnon diffraction,  is similar to
the Schrodinger equation for electron band structure.
The only difference is that the Schrodinger equation contains the electron 
energy,
while Eq.(\ref{eqphinew}) contains $\omega^2$.
Solution of Eq.(\ref{eqphinew}) is obvious from this analogy,
\begin{eqnarray}
\label{phinew}
(\Omega_{\bm q}^{(in)})^2&=&\frac{1}{2} [(\omega_{\bm q}^{(in)})^2+(\omega_{\bm q\pm2\bm Q}^{(in)})^2]\nonumber\\
&\pm&\sqrt{\frac{1}{4} [(\omega_{\bm q}^{(in)})^2-(\omega_{\bm q\pm2\bm Q}^{(in)})^2 ]^2
+M_{\bm q}^2}\, .
\end{eqnarray}
The sign $\pm$ before the square root and the sign $\pm$
in $\bm q\pm2\bm Q$ depend on the momentum $q_b$. The choice of the signs
must correspond to the standard band theory convention.
The mixing matrix element $M_{\bm q}$ 
is different for $q_b \approx Q$ and for $q_b \approx \pi - Q$.
How to find values of the matrix element?
Let us, for example, take ${\bm q}=(q_a,Q,q_c)$. 
Here the solution of Eq.(\ref{eqphinew}) must be of the following form,
$\varphi \propto e^{iq_aa+iq_cc}\psi_b$, where
$\psi_b=\cos(Qb)$ or $\psi_b=\sin(Qb)$.
Substitution of these two solutions in Eq.(\ref{eqphinew}) allows one to find
corresponding frequency $(\Omega_{\bm q}^{(in)})^2$.
On the other hand, according to (\ref{phinew}) the frequencies are
$(\Omega_{\bm q}^{(in)})^2=(\omega_{\bm q}^{(in)})^2\pm M_{\bm q}$.
Comparing we find  value of the matrix element.
This calculation gives the following results
\begin{eqnarray}
\label{M12}
&&q_b \approx Q : \\ 
&&M_{\bm q}=4\Lambda J_1 S^2\left\{\frac{3}{2}+\frac{R(q_a,0,q_c)-R(q_a,2Q,q_c)}
{R(3\bm Q)-R(\bm Q)}\right\}\nonumber\\
&&q_b \approx \pi-Q : \nonumber\\ 
&&M_{\bm q}=4\Lambda J_1 S^2\left\{\frac{3}{2}+\frac{R(q_a,\pi,q_c)-R(q_a,\pi-2Q,q_c)}
{R(3\bm Q)-R(\bm Q)}\right\}\nonumber
\end{eqnarray}

\begin{figure}[h!]
\includegraphics[width=0.25\textwidth,clip]{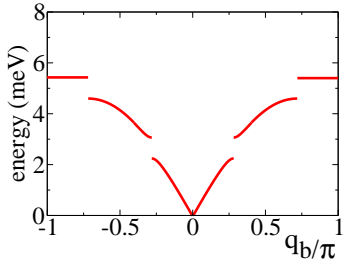}
\caption{In-plane magnon dispersion with account of the crystal field 
anisotropy (\ref{db}). The dispersion is shown along the
${\bm q}=(0,q_b,0)$ direction.
} 
\label{F5a}
\end{figure}

\begin{figure}[h!]
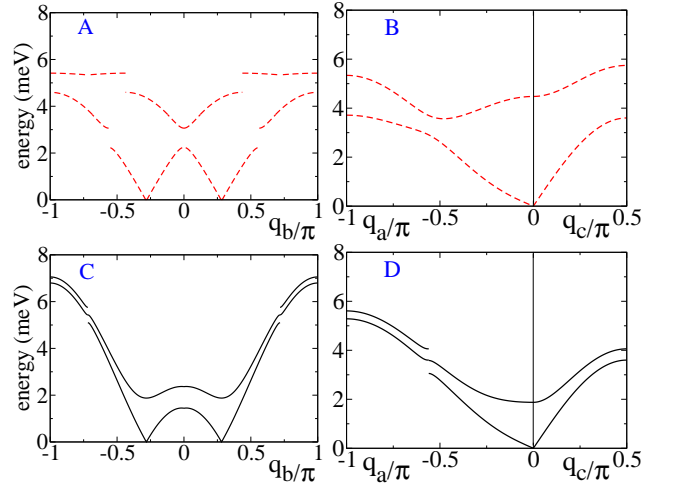

\includegraphics[width=0.23\textwidth,clip]{Fig5bI.eps}
\includegraphics[width=0.23\textwidth,clip]{Fig5cI.eps}
\includegraphics[width=0.23\textwidth,clip]{Fig5b.eps}
\includegraphics[width=0.23\textwidth,clip]{Fig5c.eps}
\caption{Magnon dispersions (as they seen in neutron scattering)
with account of the crystal field anisotropy (\ref{db}).
Panels A and B: branches of the in-plane dispersion 
$\Omega^{in}({\bm q}\pm {\bm Q})$.
Panels C and D: branches of the out-of-plane dispersion 
$\Omega^{out1}({\bm q})$, $\Omega^{out2}({\bm q})$.
Panels A and C correspond to {\bf q} along {\bf b}, ${\bm q}=(0,q_b,0)$.
Panels B and D correspond to ${\bm q}={\bm Q}+\delta{\bm q}$, where
$\delta{\bm q}$ is directed along {\bf a} and {\bf c}.
} 
\label{F5}
\end{figure}

The in-plane dispersion $\Omega^{(in)}_{\bm q}$ for 
${\bm q}=(0,q_b,0)$ is shown in Fig.\ref{F5a}. 
Discontinuities of the dispersion due to diffraction of magnons
from the static spin spiral are clearly seen.
We already pointed out that the in-plane excitation  cannot be seen directly 
in neutron scattering since the  corresponding n-field (\ref{nin}) contains 
an additional oscillating factor $\cos(\bm Q\cdot\bm r)$ or $\sin(\bm Q\cdot\bm r)$. Therefore in a scattering  measurement the in-plane mode is seen as two 
shifted branches 
$\Omega^{in}({\bm q}\pm{\bm Q})$ with half intensity each.
These branches for three different momentum directions
are shown by red dashed lines in  Fig.\ref{F5}, panels A and B.
\\

{\underline {\large Out-of-plane excitations.}}\\
There are two anisotropy induced effects on the
out-of-plane excitations, (i) opening of the gap at zero
frequency, (ii) discontinuity of the dispersion due to diffraction 
of magnons from the static spin spiral.
Without an anisotropy there are two out-of-plane Goldstone modes
with $\bm q=\pm{\bm Q}$ corresponding to linear
combinations of rotations around ${\vec e}_1$ and ${\vec e}_2$,
see Fig.\ref{F4}C, black solid line.
The anisotropy (\ref{db}) does not respect rotations around ${\vec e}_1$,
but it does respect rotations around ${\vec e}_2={\vec e}_b$.
Therefore,  we expect one gapless and one gapped out-of-plane mode.

For out-of plane fluctuations we have
\begin{eqnarray}
\label{noutnew}
{\vec n}=\sqrt{1-h^2}{\vec N}_0+h {\vec e}_3 \,,
\end{eqnarray}
and the corresponding Euler-Lagrange equation is
\begin{eqnarray}
\label{noutnew222}
&&-\chi_{\perp}{\ddot h}
+2S^2[-h{\vec N}_0 R(\bm p){\vec N}_0+{\vec e}_3 R(\bm p){\vec e}_3h]\nonumber\\
&&-S^2\Lambda[1-\cos(2\bm Q\cdot\bm r+2\varphi_{st})]h =0 \,.
\end{eqnarray}
Expanding this equation up to the first order in  $\Lambda$ we get
\begin{eqnarray}
\label{noutnew1}
&&-\chi_{\perp}{\ddot h}
+2S^2[R(\bm p)- R(\bm Q)]h\nonumber\\
&&=S^2\Lambda[1-2\cos(2\bm Q\cdot\bm r)]h \,.
\end{eqnarray}
It is easy to check that at $\bm q=\bm Q$ this equation has
a gapless solution
\begin{eqnarray}
\label{noutnew2}
\Omega_{\bm Q}^{(out1)}=0,\ \ \  h\propto\cos(\bm Q\cdot\bm r). 
 \end{eqnarray}
and a gapped solution
\begin{eqnarray}
\label{noutnew2a}
\Omega_{\bm Q}^{(out2)}=\sqrt{8S^2J_1\Lambda}, \ \ \  
h\propto\sin(\bm Q\cdot\bm r).
 \end{eqnarray}
In $\cos(\bm Q\cdot\bm r)$ and $\sin(\bm Q\cdot\bm r)$ solutions
we neglect higher harmonics terms which have small amplitudes
$\sim \Lambda/J_1$.
Thus, the spectrum near its minimum agrees with our expectations.

Eq.(\ref{noutnew1})  contains the effective scattering ``potential'' 
with momentum  $\Delta {\bm q}=2{\bm Q}$. Hence there must be
a discontinuity of the spectrum at $q_b=\pm(\pi-Q)$.
Similarly to (\ref{phinew}), the spectrum in the vicinity of this
momentum is
\begin{eqnarray}
\label{outnew}
&&(\Omega_{\bm q}^{(out)})^2\approx\frac{1}{2} [(\omega_{\bm q}^{(out)})^2+(\omega_{\bm q+2\bm Q}^{(out)})^2]
\nonumber\\
&&\pm\sqrt{\frac{1}{4} [(\omega_{\bm q}^{(out)})^2-(\omega_{\bm q+2\bm Q}^{(out)})^2 ]^2+(4J_1S^2\Lambda)^2}\,.
\end{eqnarray}
The out-of-plane excitations are seen
in inelastic neutron scattering as they are. Dispersions
$\Omega_{\bm q}^{(out1)}$ and $\Omega_{\bm q}^{(out2)}$
 along three different directions are plotted in Fig.\ref{F5}, panels C and D,
by black solid lines. 

\section{Excitation spectra with account of both the crystal field 
anisotropy and the Dzyaloshinski-Moriya anisotropy}
The effective  Dzyaloshinski-Moriya (DM)  interaction between the ferroelectric
polarization ${\vec P}$ and spins is of the following form 
\cite{Katsura2005,Mostovoy2006}
\begin{equation}
\label{DM}
H_{DM}\propto {\vec P}\cdot\left[{\vec e}_{12}\times
\left[{\vec S}_1\times{\vec S}_2\right]\right] \ ,
\end{equation}
where ${\vec S}_1$ and ${\vec S}_2$ are spins at nearest sites
and ${\vec e}_{12}$ is a unit vector directed from the site 1 to the
site 2. 
Here we consider the case of zero external magnetic 
field when  the polarization ${\vec P}$ is directed 
along the c-axis~\cite{Kimura2003}. The vector ${\vec e}_{12}$
is directed along the b-axis and hence the interaction (\ref{DM}) put the 
spin spiral in the bc-plane. 
\begin{equation}
\label{ebc}
{\bm e}_2={\bm e}_b\ , \ \ \  {\bm e}_1 ={\bm e}_c\ .
\end{equation}
Eq. (\ref{DM}) can be rewritten in terms
of the unit vector ${\vec n}$ describing the magnetization
staggered in the c-direction,
\begin{equation}
\label{DM1}
H_{DM}={\cal D} S^2[\vec{n}\times \nabla_b\vec{n}]_a \to const+
{\cal D} Q S^2 n_a^2 \ , 
\end{equation} 
where ${\cal D} >0$ is the constant of the DM interaction.
So, in these notations the DM interaction is
equivalent to the crystal field anisotropy with the coefficient 
in the effective crystal field 
proportional to the wave vector of the spin spiral.
The coefficient ${\cal D}$ is related to the ferroelectric polarization and
therefore it is strongly temperature dependent. In particular
${\cal D}=0$ in the spin stripe phase at $T > T_S$.
However, here we consider the system deep in the spin spiral phase, 
$T < T_S$, and for numerical estimates we use
\begin{equation}
\label{gaa}
{\cal D}=0.20 \ {\mbox {meV}}\ ,
\end{equation}
which results in spectra approximately consistent with the neutron scattering 
data~\cite{Senff2008}.\\

\begin{figure}[h!]
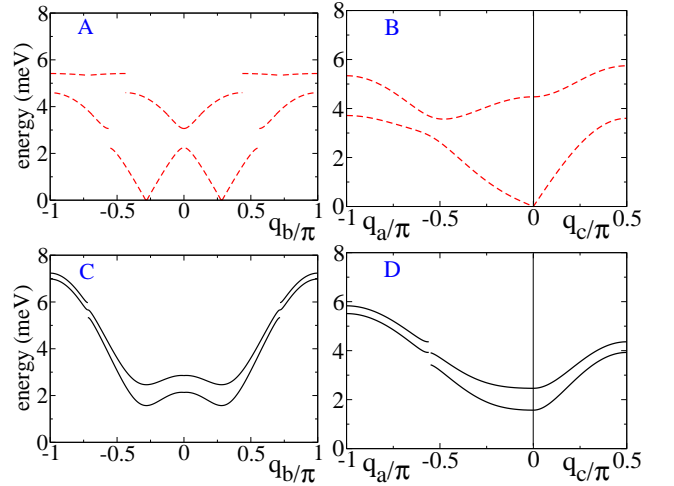

\includegraphics[width=0.23\textwidth,clip]{Fig5bI.eps}
\includegraphics[width=0.23\textwidth,clip]{Fig5cI.eps}
\includegraphics[width=0.23\textwidth,clip]{Fig8b.eps}
\includegraphics[width=0.23\textwidth,clip]{Fig8c.eps}
\caption{Magnon dispersions (as they seen in neutron scattering)
with account of both the crystal field anisotropy (\ref{db}) and the
Dzyaloshinski-Moriya anisotropy (\ref{DM}),(\ref{DM1}).
Panels A and B: branches of the in-plane dispersion 
$\Omega^{in}({\bm q}\pm {\bm Q})$.
Panels C and D: branches of the out-of-plane dispersion 
$\Omega^{out1}({\bm q})$, $\Omega^{out2}({\bm q})$.
Panels A and C correspond to {\bf q} along {\bf b}, ${\bm q}=(0,q_b,0)$.
Panels B and D correspond to ${\bm q}={\bm Q}+\delta{\bm q}$, where
$\delta{\bm q}$ is directed along {\bf a} and {\bf c}.
} 
\label{F8}
\end{figure}

{\underline {\large In-plane excitations.}}\\
The DM anisotropy obviously does not influence the in-plane spin
fluctuations. Therefore the in-plane excitation spectra derived
in Section IV are fully valid in this case. 
In Fig.\ref{F8} we present magnetic excitation spectra
with account of both the crystal field anisotropy  and the
Dzyaloshinski-Moriya anisotropy.
Panels A and B in Fig.\ref{F8} are identical to that in Fig.\ref{F5}.
\\

{\underline {\large Out-of-plane excitations.}}\\
We remind that even with the crystal field anisotropy
but without of  the DM anisotropy one of the out-of-plane
excitations modes remains gapless, see panels C and D in 
Fig.\ref{F5}.
The most notable effect of the DM anisotropy is opening of
a gap in the remaining gapless mode. With account of 
the anisotropy
Eqs. (\ref{noutnew1}), (\ref{noutnew2}), and {\ref{noutnew2a}) are
modified as
\begin{eqnarray}
\label{noutnew1DM}
&&-\chi_{\perp}{\ddot h}
+2S^2[R(\bm p)- R(\bm Q)]h\nonumber\\
&&=S^2\left\{\Lambda[1-2\cos(2\bm Q\cdot\bm r)]+ {\cal D}Q\right\}h \,.
\end{eqnarray}
At $\bm q=\bm Q$ this equation has two gapped solutions
\begin{eqnarray}
&&\Omega_{\bm Q}^{(out1)}\approx\sqrt{4S^2J_1{\cal D} Q},
\ \   h\propto\cos(\bm Q\cdot\bm r)
\\ \label{noutnew2aDM}
&&\Omega_{\bm Q}^{(out2)}\approx \sqrt{4S^2J_1(2\Lambda+{\cal D} Q)},
\ \    h\propto\sin(\bm Q\cdot\bm r)\ . \nonumber
 \end{eqnarray}
In $\cos(\bm Q\cdot\bm r)$ and $\sin(\bm Q\cdot\bm r)$ solutions
we neglect higher harmonics terms which have small amplitudes
$\sim \Lambda/J_1$.

Similarly to Eq.(\ref{noutnew1}), 
Eq.(\ref{noutnew1DM})  contains the effective scattering ``potential'' 
with momentum  $\Delta {\bm q}=2{\bm Q}$. Hence there must be
a discontinuity of the spectrum at $q_b=\pm(\pi-Q)$.
Similarly to (\ref{outnew}), the spectrum in the vicinity of this
momentum is
\begin{eqnarray}
\label{outnewDM}
&&(\Omega_{\bm q}^{(out)})^2\approx\frac{1}{2} [(\omega_{\bm q}^{(out)})^2+(\omega_{\bm q+2\bm Q}^{(out)})^2]
\\
&&\pm\sqrt{\frac{1}{4} [(\omega_{\bm q}^{(out)})^2-(\omega_{\bm q+2\bm Q}^{(out)})^2 ]^2+(4J_1S^2\Lambda)^2}\,.
\end{eqnarray}
The out-of-plane excitations are seen
in inelastic neutron scattering as they are. Dispersions
$\Omega_{\bm q}^{(out1)}$ and $\Omega_{\bm q}^{(out2)}$
 along three different directions are plotted in Fig.\ref{F8}, panels C and D,
by black solid lines. 

\section{Conclusions}
We have calculated spectra of magnetic excitations in the spin spiral state
of perovskite manganates TbMnO$_3$ and DyMnO$_3$.
As starting point we use the frustrated Heisenberg Hamiltonian 
$H+\delta H$ suggested in Refs.~\cite{Moussa1996,Hirota1996,Senff2008}
and determined by Eqs. (\ref{h1}), (\ref{h2}).
We also account for the crystal field anisotropy (\ref{db}) and 
the Dzyaloshinski-Moriya anisotropy, (\ref{DM}),(\ref{DM1}).
In the present work we do not consider a relaxation, hence a line broadening
is not included in the analysis.

To simplify calculations and to get a physical insight in the structure of
magnetic excitations we employ a  $\sigma$-model like field theory.
At small momenta in the region
of the most important and complex  incommensurate physic,
the field theory  is fully equivalent to the Heisenberg model.
On the hand,
close to the boundary of magnetic Brillouin zone the field theory
underestimates the magnon frequency by about 20\% compared to the
Heisenberg model.
Values of parameters which reproduce the measured 
dispersion in TbMnO$_3$, Ref.~\cite{Senff2008}, are listed in Eqs. 
(\ref{parameters}), (\ref{gbb}), and (\ref{gaa}).
Exchange integrals in Eq. (\ref{parameters}) are consistent with that
in Ref.~\cite{Senff2008} with account of different definitions (factor 2).

There are in-plane (spin oscillates in the plane of the spin spiral)
and out-of-plane excitations (spin oscillates perpendicular to the plane 
of the spin spiral). Dispersions of the in-plane and the out-of-plane 
excitations (as they are seen in neutron scattering) without account
of the crystal field and Dzyaloshinski-Moriya anisotropies are presented
in Fig.~\ref{F4}. All the dispersions are Goldstone ones,
the energy is zero at the wave vector equal to the wave vector of the spin
spiral.

Account of the crystal field anisotropy leads to the two effects
(i) opening of the gap in one of the Goldstone modes,
(ii) discontinuity of the dispersion due to diffraction 
of magnons from the static spin spiral.
Dispersions of the in-plane and the out-of-plane 
excitations (as they are seen in neutron scattering) with account
of the crystal field anisotropy but without account of the
Dzyaloshinski-Moriya interaction are presented
in Fig.~\ref{F5}.

Further account of the Dzyaloshinski-Moriya interaction opens gap in
both out-of-plane modes. As expected, the in-plane sliding mode remains gapless
in spite of the anisotropies.
Dispersions of the in-plane and the out-of-plane 
excitations (as they are seen in neutron scattering) with account
of both  the crystal field anisotropy and the
Dzyaloshinski-Moriya interaction are presented in Fig.~\ref{F8}.
These curves agree pretty well with experimental data from 
Ref.~\cite{Senff2008} Note, there are three different dispersion 
curves in the low energy ($\omega  < 3$ meV) region.

\acknowledgments
We thank Clemens Ulrich, Narendirakumar Narayanan, and Maxim Mostovoy
for important stimulating discussions.
A.~I.~M. gratefully acknowledges the Faculty of Science and the 
School of Physics at the University of New
South Wales for warm hospitality during his visit.
O.~P.~S. gratefully acknowledges  Yukawa Institute for Theoretical Physics
for warm hospitality during work on this project.


\begin{thebibliography}{99}
\bibitem{com1} The trick with extension of the effective field theory
up to the boundary of the antiferromagnetic Brillouin zone is only
possible if the coupling is antiferromagnetic only along one direction.
\end{thebibliography}
\end{document}